\documentclass[fleqn,10pt]{wlscirep}
\usepackage{siunitx}             
\usepackage{graphicx}            
\usepackage{xcolor}              
\usepackage{subfigure}           
\usepackage{amsmath}             
\usepackage{epstopdf}

\begin{document}
%
%



\title{A misaligned magneto-optical trap to enable miniaturized atom chip systems}

\author[1,2,$\dagger$]{Ritayan Roy}
\author[1]{Jo Rushton}
\author[1]{Andrei Dragomir}
\author[1]{Matthew Aldous}
\author[1,*]{Matt Himsworth}

\affil[1]{School of Physics and Astronomy, University of Southampton, Highfield, Southampton, SO17 1BJ, United Kingdom}
\affil[2]{Present address: School of Mathematical and Physical Sciences, University of Sussex, Falmer Campus, Brighton, BN1 9QH, United Kingdom}
\affil[*]{m.d.himsworth@soton.ac.uk}
\affil[$\dagger$]{ritayan.roy@u.nus.edu}

%
%

\begin{abstract}
We describe the application of displaced, or misaligned, beams in a mirror-based magneto-optical trap\,(MOT) to enable portable and miniaturized atom chip experiments, where optical access is limited to a single window. Two different geometries of beam displacement are investigated: a variation on the well-known `vortex-MOT', and the other a novel `hybrid-MOT' combining Zeeman-shifted and purely optical scattering force components. The beam geometry is obtained similar to the mirror-MOT, using a planar mirror surface but with a different magnetic field geometry more suited to planar systems. Using these techniques, we have trapped around 6$\times 10^6$ and 26$\times 10^6$atoms of $^{85}$Rb in the vortex-MOT and hybrid-MOT respectively. For the vortex-MOT the atoms are directly cooled well below the Doppler temperature without any additional sub-Doppler cooling stage, whereas the temperature of the hybrid-MOT has been measured slightly above the Doppler temperature limit. In both cases the attained lower temperature ensures the quantum behaviour of the trapped atoms required for the applications of portable quantum sensors and many others. 
\end{abstract}

\flushbottom
\maketitle
\thispagestyle{empty}

\section*{Introduction}

Progress in science and technology over the last decades has shown that miniaturization and integration can lead to robust applications of fundamental physics, be it the miniaturization of electronics by integrated circuits or in optics in terms of micro-optical devices and sensors. In atomic physics, atom chip experiments based on neutral atoms or ions are starting to be realized in similar scalable quantum-optical systems. An atom chip, at its most basic, is a substrate with microfabricated conductors which produce magnetic and electric fields that can be used to trap and manipulate atoms and ions. Atom chips enable highly sophisticated experiments to be condensed into areas on the order of a few square centimetres and readily lend themselves to the miniaturization and integration of cold atom systems for practical applications beyond the laboratory.

The concept of creation of magnetic traps using micro-structured chips is proposed by Weinstein $et\ al.$ in 1995~\cite{Weinstein_1995}. There are development with discrete wires \cite{Schmiedmayer_1995, Haase_2001, Fortagh_1998, Denschlag_1999, Vuletic_1998} and permanent magnets~\cite{Roach_1995, Sidorov_1996, Hughes_1997, Saba_1999}, but the first successful realization of a trap on an atom chip is by Reichel $et\ al.$ \cite{Reichel_1999} and Folman $et\ al.$~\cite{ Folman_2000} using a technique called the mirror-MOT. The beam geometry of the standard mirror-MOT is incompatible with microfabricated vacuum cells with limited optical access \cite{IntegratedAtomChipFeasibility, salim2011}. There are various schemes to create a MOT within small volumes \cite{Lee96PyramidMOT, SiPyramidMOT, TetraMOT, GMOT}, however they have complex and expensive microfabrication procedures and several of these designs are not easily compatible with planar atom chip structures.

Recently, to solve the above limitations, a novel switching-MOT technique (S-MOT) \cite{rushton2016} is reported which relies on the synchronized dynamic switching of magnetic and optical fields. In this scheme, the magnetic fields are produced solely from planar conductors, and all beams are incident on the mirror at \SI{45}{\degree} thus only requiring a single viewport.  To simplify the S-MOT technique further, we have investigated how it would be possible to make a MOT without the switching of the optical and magnetic fields whilst retaining the same advantages. Here, we demonstrate two MOT techniques based on a DC magnetic field with displaced MOT beams which are referred to as the \emph{vortex-MOT} and \emph{hybrid-MOT}. The vortex-MOT, which is formed by displacing the MOT beams in a vortex pattern (see Figs. \ref{experimental_setup}(d) and \ref{vsim}), is in a similar vein to an earlier demonstration \cite{Shimizu:91} but adopted for a single optical window. In contrast, the novel hybrid-MOT (Figs. \ref{experimental_setup}(d) and \ref{hsim}) technique is created by an elegant balancing between the trapping forces created via pure optical scattering and Zeeman-shifted resonances.


%

\section*{Theory}
\begin{figure}[tb]
\centering
\subfigure[]{\includegraphics[width=0.3\textwidth]{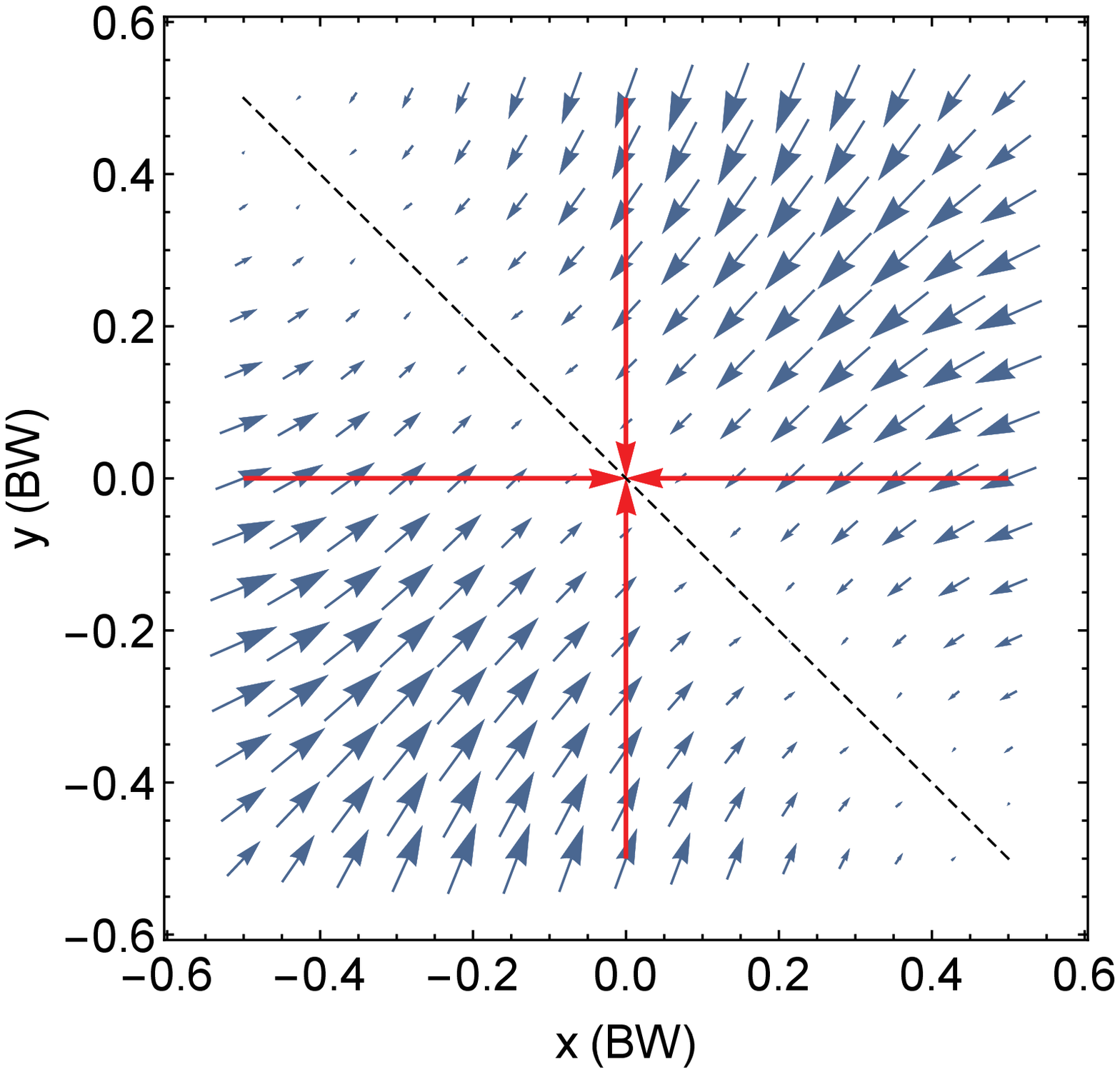}\label{rsim}}
\subfigure[]{\includegraphics[width=0.3\textwidth]{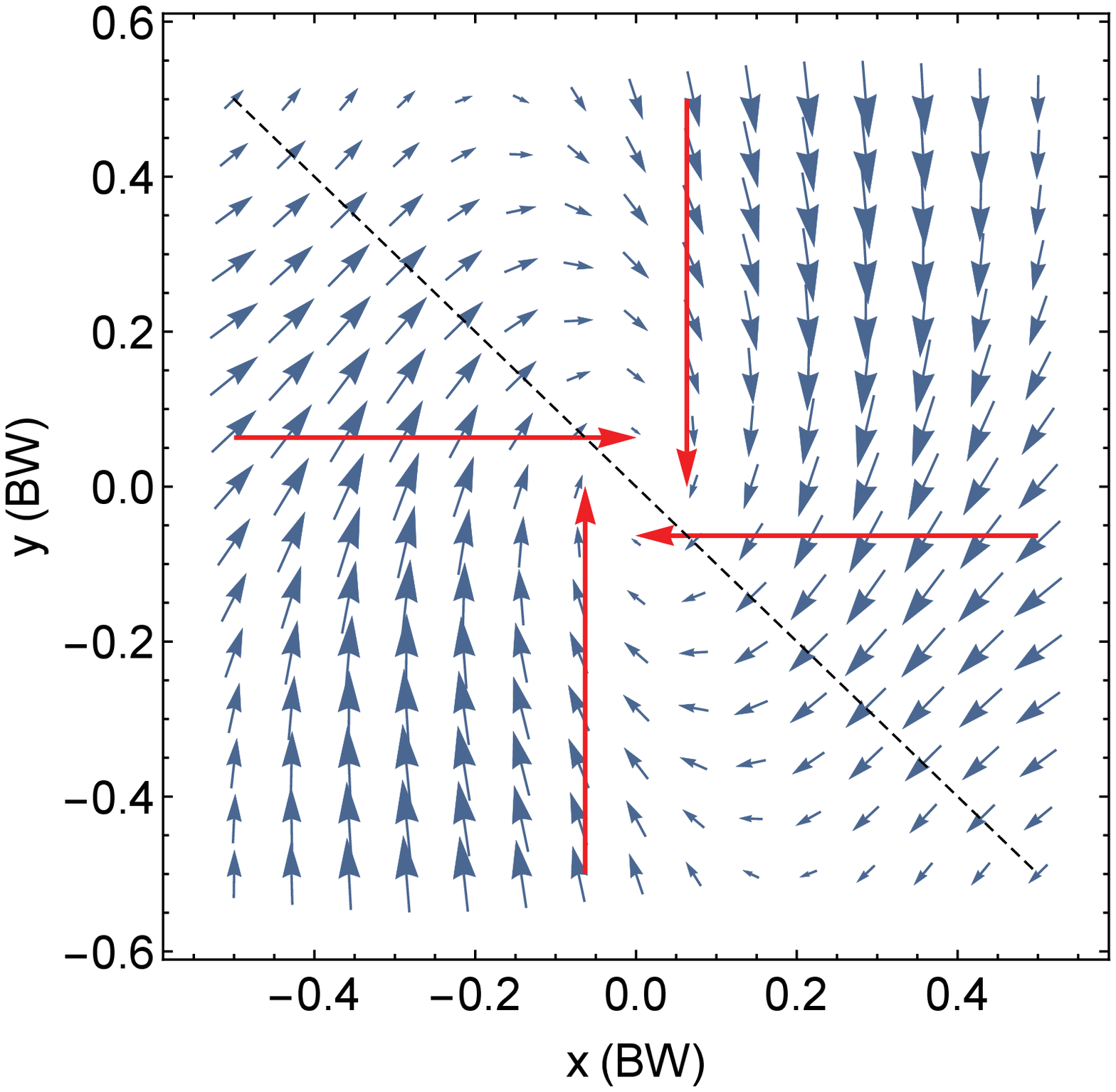}\label{vsim}}
\subfigure[]{\includegraphics[width=0.3\textwidth]{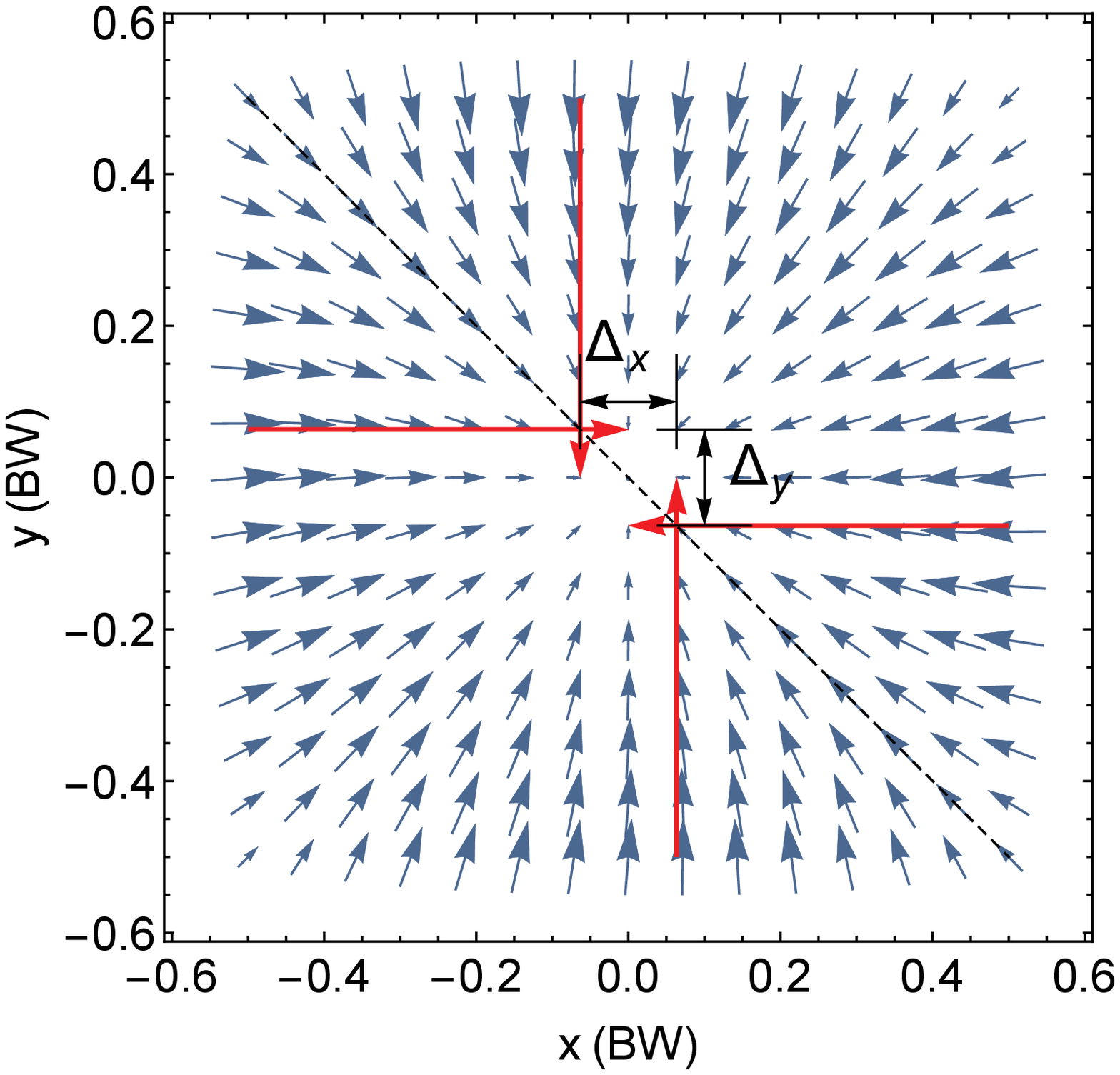}\label{hsim}}

 \caption{Analytical simulations of the restoring force vectors (blue arrows) on a stationary atom in the $xy$ plane. The red arrows indicate the direction of the beams incident at \SI{45}{\degree} along the $z$ axis out of the page. The tips of the red arrows signify the point of reflection on the mirror surface. In a) there is no misalignment resulting in no restoring force along the diagonal highlighted with a dashed line. By misaligning the beams shown in b) the beams produce a vortex-like restoring force, whereas swapping the beams propagating in the $y$ direction results in the `hybrid' restoring force. The separations between along the $x$ and $y$ axes the beams are $\Delta_x$ and $\Delta_y$, respectively and always symmetrically around the magnetic zero point at $x=y=0$. }
\label{simulation}
\end{figure}

The S-MOT is based around a quasi-quadrupole magnetic field produced by two orthogonal pairs of separated conducting wires. The aim is to find an alternative planar geometry to produce the correct field gradients, in much the same manner as in Wildermuth \emph{et al}. \cite{wildermuth2004optimized}, but avoiding any external bias fields. The lack of rotation symmetry of this geometry resulted in a axis over which no trapping occurs, as shown in Fig. \ref{rsim}, and so our effort with the S-MOT, and this report, is to explore methods to close off this loss channel. 

During experimentation of the S-MOT, it is noticed on several occasions that a dense cloud of atoms would form when all optical and magnetic fields are static and the beams are not perfectly aligned. Obtaining the cloud is surprisingly trivial as it would form quite readily, and significant effort is employed to avoid it whilst demonstrating the separate S-MOT mechanism. The range of operational parameters over which the static trap worked (alignment, polarization, beam intensity, detuning) ruled out a coherent argument for a single mechanism, but we believe most can be explained via two methods reported here. These trap geometries differ only via a small displacement on one pair of beams (see Fig. \ref{simulation}), however their principles of operation have a far greater disparity. We note that for both mechanisms, as well as the S-MOT, trapping and cooling in the $z$ direction, orthogonal to the mirror, occurs via the usual MOT mechanism, and we are only concerned here with obtaining trapping in the $xy$ plane above the mirror.

\subsection*{Vortex-MOT}
In this geometry, shown in Fig. \ref{vsim}, each counter-propagating pair of beams are displaced by the same distance from the incident plane, but in orthogonal directions. For example, the beam propagating in the $\left[+x, +z\right]$ plane is displaced by a small distance along the $+y$ axis, whereas the beam propagating along the $\left[-x, +z\right]$ plane is displaced along $-y$. The resulting restoring force is not directed toward the trap center but instead follows a curl around the center - like a vortex - producing a torque on the atomic cloud. Traps operating on a similar manner, albeit not involving a mirror, have been demonstrated several times in the literature, and this mechanism is responsible for `race track' patterns occasionally formed during the alignment of standard MOTs \cite{Sesko91, bagnato93}. We will not discuss the theory here, as this has been thoroughly explored in the literature \cite{Guedes94, Arnold00,Felinto00}. For trap parameters in which the cooling and restoring forces are greater than the vortex torque, the atoms will spiral in toward the trap center forming a spherical cloud. This is typically achieved when the beam separation, $\Delta$, is less than the beamwidth, $\rho$, and is the parameter space this report explores. For weaker traps, and/or larger separations the cloud will begin for form rings around a small central cloud, or solely ring-shaped clouds.  
    
\subsection*{Hybrid-MOT} 
The hybrid mechanism is so-called as it encompasses both the usual Zeeman-shift dependent restoring force of the MOT (which we refer to as the `Zeeman' force) along one direction, and the purely optical scattering force along the orthogonal direction (referred to as the `Optical' force). By displacing the beams by a small amount symmetrically away from the trap center, an atom traversing the beam overlap volume experiences a variation in intensity between the two beams. The Gaussian beam-shape results in a scattering minimum at the trap centre and results in a fully restoring force to the center, as shown in Fig. \ref{hsim}. This mechanism is described in the earlier discussions within the literature of trapping via the scattering force \cite{Pritchard86} and is related to the `two-beams-trap' in which both the intensity and wavevector varies across the trap. To achieve a compact and spherical cloud of trapped atoms, the Optical and Zeeman force gradients must be balanced at the trap center. To find the optimum separation of the beams we simply model the Optical force as two equal and opposite Gaussian beams, with $1/e^2$ diameter, $\rho$, separated by $\Delta/2$ from the trap center. Here, we shall use the $\left[u,v\right]$ coordinate system to define the orthogonal axes in which the Zeeman and Optical force act independently to simplify the model. In relation to Fig. \ref{hsim}, the $u$ axis corresponds to the diagonal marked by the dashed line where there is no magnetic field gradient.    
\begin{equation}
\label{opticalforce}
F_{s}(u)=\frac{\hbar k \Gamma S}{2\left(1+\left(\frac{2\delta}{\Gamma}\right)^{2}\right)}\left(\exp\left[\frac{-\left(u+\Delta/2\right)^2}{\left(2\left(\rho/4\right)^2\right)}\right]-\exp\left[\frac{-\left(u-\Delta/2\right)^2}{\left(2\left(\rho/4\right)^2\right)}\right]\right)
\end{equation}
where $k$ is the beam wavevector, $\Gamma$ is the natural linewidth, $\delta$ is the detuning from resonance (assumed negative for cooling), and $S$ is the saturation power.
This equation can be simplified to a linear function in $u$,
\begin{equation}
\label{opticalforcesim}
F_{s}\left(u\right)=\frac{\hbar k \Gamma S \Delta \left|u\right|}{2 \left(\rho/4\right)^2 \left(1+\left(\frac{2\delta}{\Gamma}\right)^{2}\right)}+O\left(u^2\right).
\end{equation}
 The Zeeman force can be found in standard textbooks with a linear approximation of
\begin{equation}
\label{zeemanforce}
F_{z}\left(v\right)=\frac{-8 k S \delta\left|v\right|}{\Gamma \left(1+\left(\frac{2\delta}{\Gamma}\right)^{2}\right)^{2}}g\mu_{B}\left|\dfrac{dB}{dv}\right|, 
\end{equation}
where g is the Land$\acute{e}$ g-factor, $\mu_B$ is the Bohr magneton, and $\dfrac{dB}{dv}$ is the magnetic field gradient. We have assumed throughout that the beam intensities do not saturate the transition, so that the beams can be treated independently. By equating equation \ref{opticalforcesim} to equation \ref{zeemanforce} with $\left|u\right|=\left|v\right|$, and solving for $\Delta$ we find the optimum separation:
\begin{equation}
\label{optsep}
\Delta=\frac{-\alpha \rho^{2}\delta g \mu_{B}}{\hbar\Gamma^2\left(1+\left(\frac{2\delta}{\Gamma}\right)^{2}\right)}\left|\dfrac{dB}{dv}\right|,
\end{equation}
where $\alpha$ is a geometrical factor which takes into account the increase in beamwidth due to the incident angle and any variation in separation due to off-axis alignment. For this work in which the $\left[u,v\right]$ axes are at \SI{45}{\degree} to the $\left[x,y\right]$ axes, $\alpha=1/\sqrt{2}$, and the beamwidth along $u$ remains unaltered from $\rho$. Equation \ref{optsep} is valid for $\Delta\leq \left|0.5\,\rho\right|$, within which the Optical and Zeeman forces are approximately linear. As expected, this separation is highly dependent on beam width, as well as detuning, but independent of the beam intensity as both Optical and Zeeman forces are both mediated by photon scattering and the saturation parameter, $S$, cancels out.

An important aspect to note is the beam displacements should always be along the axis perpendicular to the beam propagation. From analytical simulations we have found that any misalignment along the wavevector axis causes an imbalance of beam intensity along the axis orthogonal to the mirror plane, $z$, thus shifting the trap center out of the beam overlap volume. The vortex-MOT is  experimentally easier to produce as the atom cloud forms with a wide range of beam misalignments, even where the $x$ and $y$ displacement are not the same. The Hybrid mechanism is far more sensitive to the beam separation. For example, in our system, the $7.9\,$mm diameter beams should be displaced by $\Delta=0.5$\,mm, which is quite challenging, with a limit of trapping at $\Delta<1\,$mm after which the Optical force overcomes the Zeeman force.      


\section*{Methods}

\begin{figure}[tb]
\includegraphics[width=0.9\textwidth]{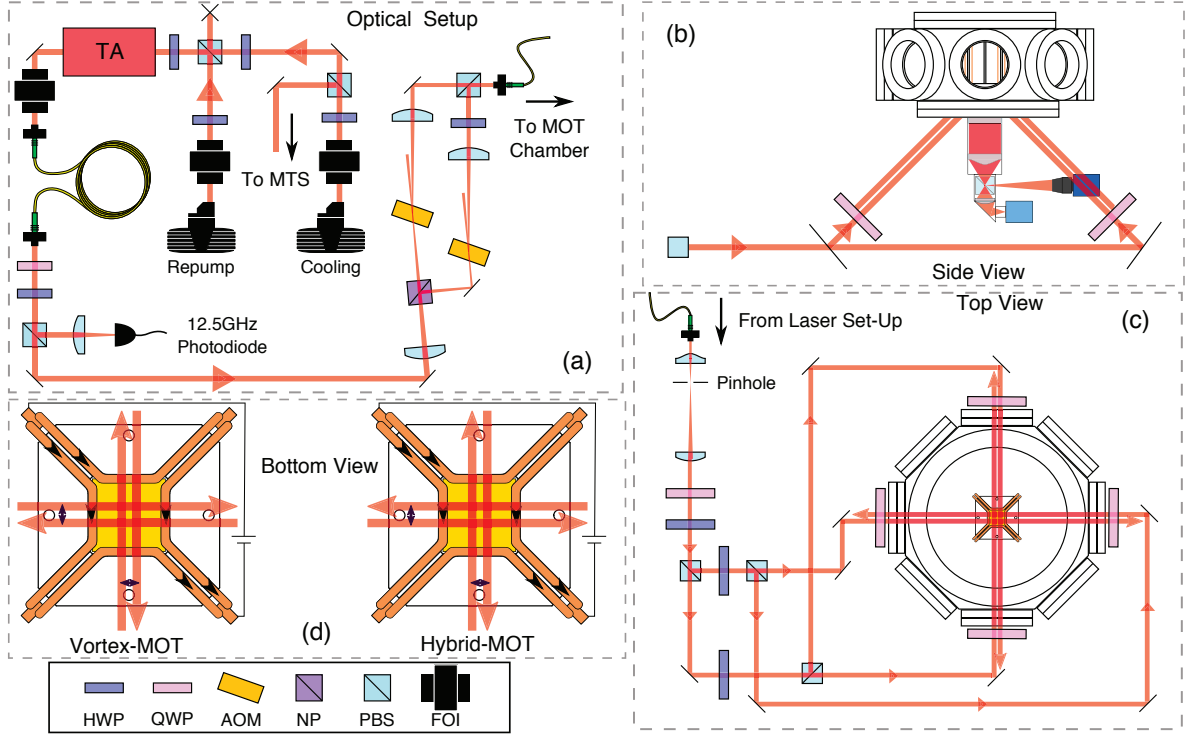}
\caption{\emph{a)} The schematic of the optical setup used to generate the cooling and repump lasers for the MOT. \emph{b)} The side view of the vacuum chamber with imaging optics. The MOT geometries enables an increased imaging capability, as a large solid angle devoid of lasers is available for detection optics. For simplicity only one pairs of beams are shown. \emph{c)} The beams are cleaned with a pinhole and separated before being directed into the vacuum chamber. \emph{d)} A diagram of the wires used to generate the quadrupole field for the vortex-MOT and hybrid-MOT are shown, in addition to the beam displacements and their directions.}
\label{experimental_setup}
\end{figure}

The laser setup for the experiment is shown in Fig. \ref{experimental_setup} \emph{(a)}. A home-made external cavity diode laser\,(ECDL) provides the cooling light, which is frequency stabilized to the $5^{2}\text{S}_{1/2}, F=3 \rightarrow{} 5^{2}\text{P}_{3/2}, F'=4$ cooling transition of ${^{85}}$Rb via modulation transfer spectroscopy (MTS) \cite{Shirley:82, Negnevitsky2013}.  The beam from this laser\,(cooling laser) is combined with another beam from a similar ECDL, the repump laser. The combined beams are used to seed a tapered amplifier (TA, m2k TA-0785-2000-DHP), whose output is cleaned by passing through a single-mode fibre. A portion of the TA's output is directed onto a fast photodiode (EOT ET-4000) and the resulting beat note is used to offset lock the repump from the cooling laser by the method described by J. Appel \emph{et al.}\cite{PhaseLock}.

The cooling laser is detuned from the ${^{85}}$Rb cooling transition using acousto-optical modulators\,(AOMs) for the requirement of the cooling and trapping of the Rb atoms. A detailed description of the optical setup is provided in the thesis of J. Rushton\cite{JRushtonThesis}. The combined cooling and repump beams are then sent to the MOT chamber using a non-polarization-maintaining optical fiber. At the other end of the fiber the beams are thoroughly cleaned with a spatial filter before being expanded to a $1/e^2$ diameter of $\rho=$\SI{7.9}{\milli\metre}. The beam is then distributed equally\,(in power) into four parts using three polarizing beam splitters and directed into the ultra-high vacuum (UHV) chamber with an incident angle of \SI{45}{\degree} to the mirror surface. The chamber has a single anti-reflection coated viewport and is maintained at a pressure of \SI{2e-9}{\milli\bar} as measured from the lifetime of its trapped atoms \cite{ArpornthipPressure}.

The quadrupole magnetic field is generated by two parallel pair of wires, where both pairs carry current in the same direction which is situated inside the vacuum chamber. Each pair produces a two-dimensional approximation to a quadrupole magnetic field and so can only confine atoms to a line equidistant between the wires. As mentioned earlier, two perpendicular pairs of wires produce the 3D quadrupole field, but the lack of rotational symmetry results in a region with no magnetic field gradient, and thus no trapping (see Fig. \ref{rsim}). 

An iris is used to reduce the beam diameter from \SI{7.9}{\milli\metre} to $0.8$\,mm only for the alignment of the displaced beams. With the aid of fluorescence and the scattering of the beams from the mirror surface, the separation of the beams are measured to an accuracy of ~$\pm 0.2$\,mm. To demonstrate the vortex-MOT, one pair of beams is displaced around $\Delta_x =1.3$\,mm from the intensity maxima where as the other pair $\Delta_y =1.0$\,mm. In the hybrid-MOT configuration, one pair of beams are separated by $\Delta_x =0.7$\,mm and the other pair by $\Delta_y =0.5$\,mm. These values are measured by analyzing fluorescence images of the beams. The beams alignments are chosen to maximise the atom number and produce the most spherical cloud of atoms, hence the slight asymmetry, but agree with the theoretical optimum displacements within their uncertainty. We suspect these differences are due to slight intensity imbalances of the beams, as well as residual magnetic fields.

The MOT being close to the viewport allows a large solid angle devoid of lasers for the fluorescence detection optics. Exploiting the favourable detection condition, as shown in the Fig. \ref{experimental_setup} \emph{(b)}, we obtain a numerical aperture\,(NA) of ${\sim}0.6$ using a aspheric condenser lens (Thorlabs ACL5040U-B). The effect of the Earth's magnetic field is reduced with 3 external nulling coils. The details about the 2D quadrupole magnetic field generation is provided in the article \cite{rushton2016} and the whole experimental setup is summarized in Fig. \ref{experimental_setup}. The quadrupole field gradient is estimated to be 9 G/cm along the $x$ and $y$ axes.

\begin{figure}[tb]
\centering
\subfigure[] {\includegraphics[width=0.49\textwidth]{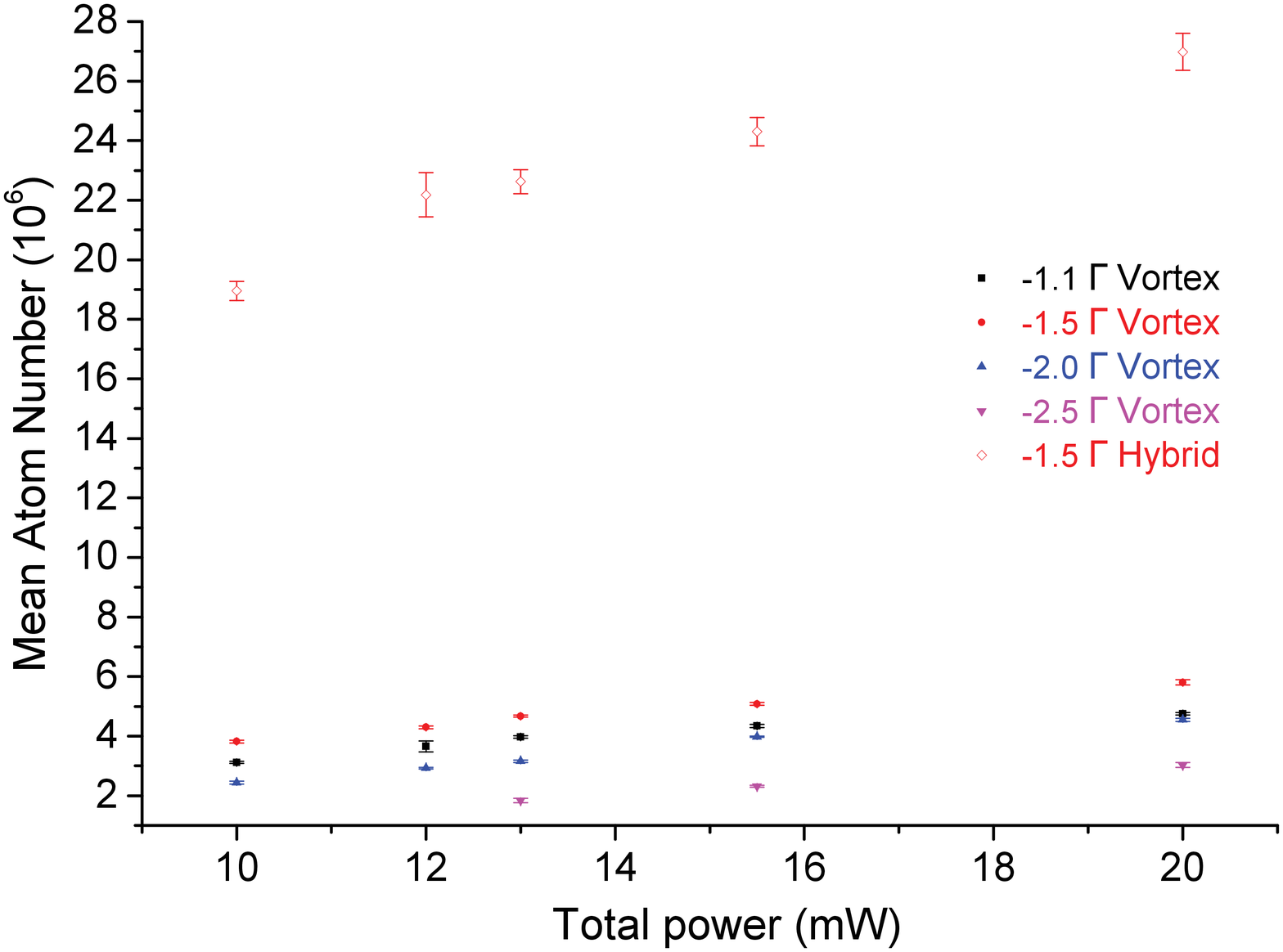}\label{AtomNo}}
\subfigure[]{\includegraphics[width=0.49\textwidth]{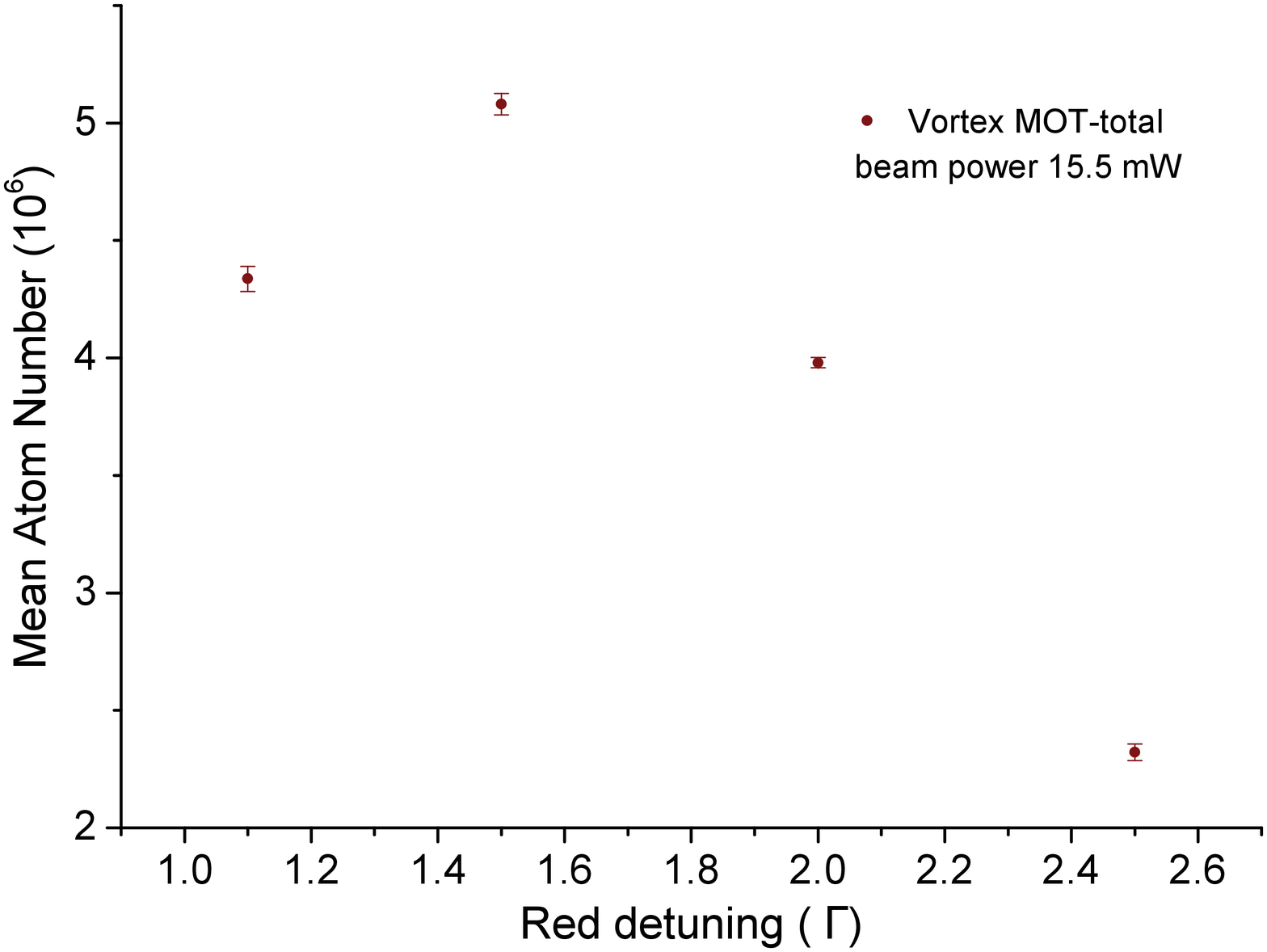}\label{AtomDet}}

 \caption{The number of trapped atoms for the vortex-MOT is measured as a function of the total MOT beam power and the red detuning of the cooling laser beams from the ${^{85}}$Rb cooling transition. For the hybrid-MOT the atom number are measured as a function of the total MOT beam power, keeping the optimum red detuning of the cooling laser beams at -1.5 $\Gamma$ from the ${^{85}}$Rb cooling transition. \emph{(a)} The mean atom number in the vortex and hybrid-MOT as a function of the MOT beam power. \emph{(b)} The mean atom number of the vortex-MOT as a function of the cooling beam red-detuning. }
\label{PhiMOT_Atom_New}
\end{figure}

\section*{Results}

\begin{figure}[tb]
\centering
\subfigure[]{\includegraphics[width=0.49\textwidth]{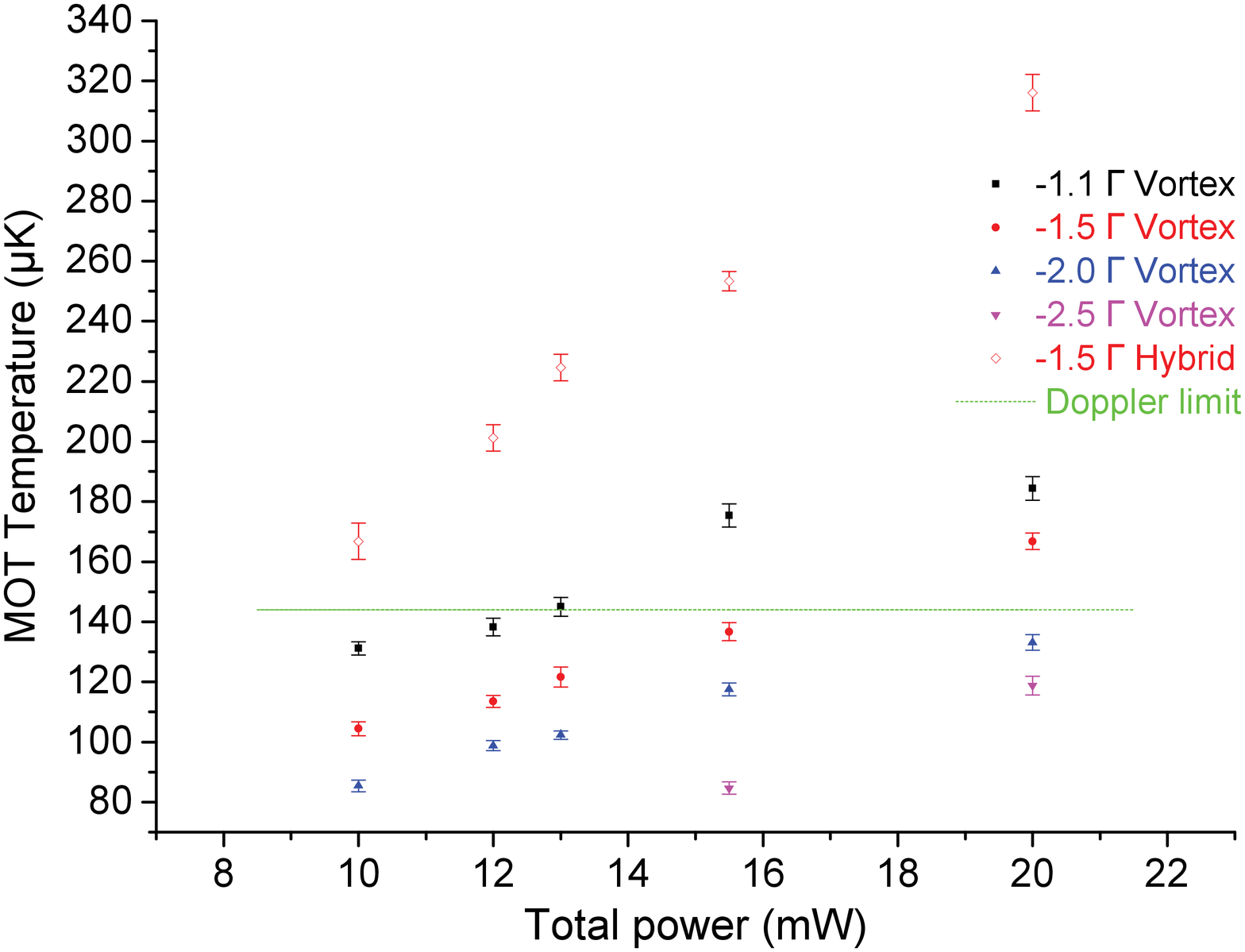}\label{Temp}}
\subfigure[]{\includegraphics[width=0.49\textwidth]{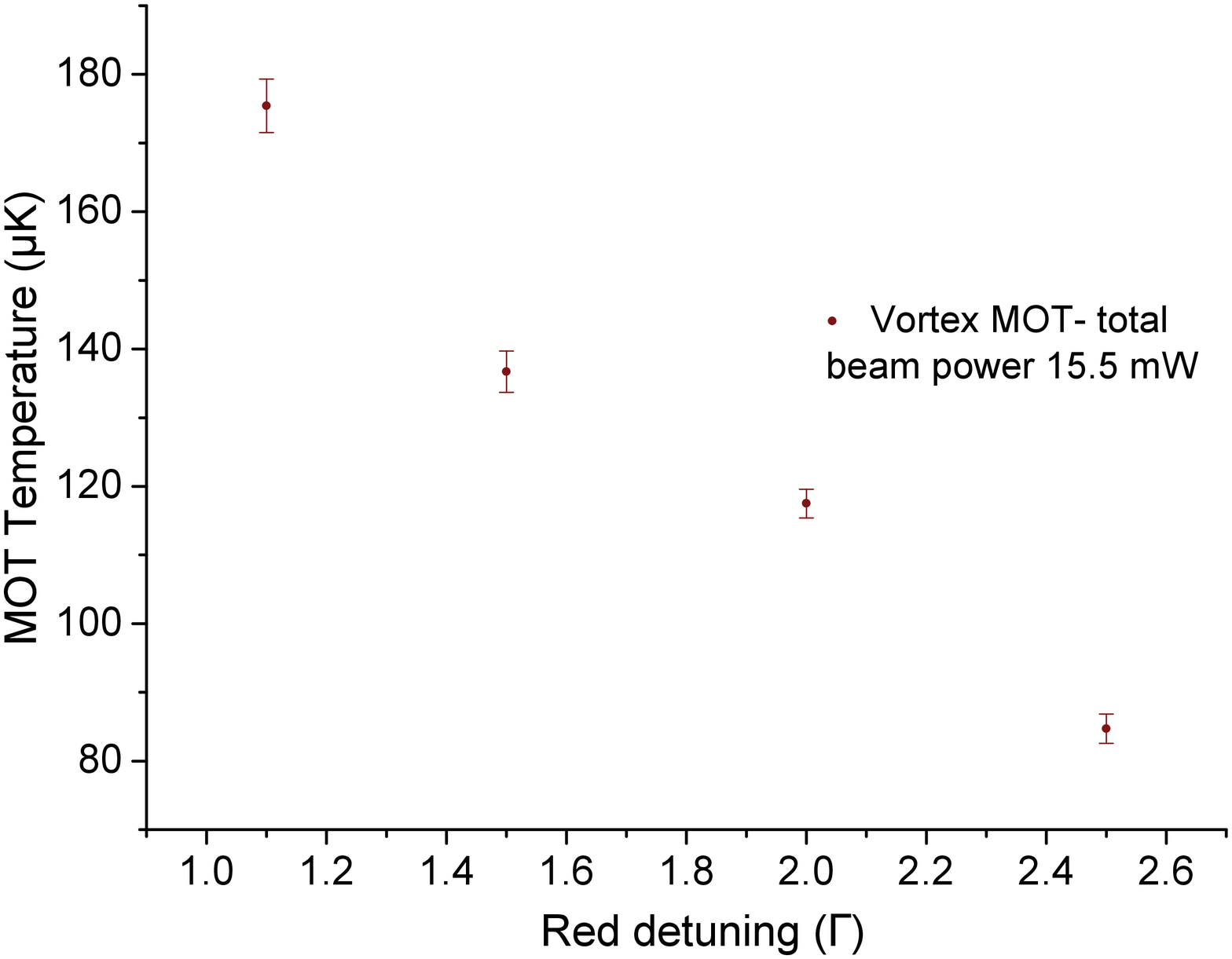}\label{TempDet}}

 \caption{The temperature of the vortex-MOT is measured as a function of the total MOT beam power and the red detuning of the cooling laser from the ${^{85}}$Rb cooling transition. For the hybrid-MOT the temperature is measured as a function of the total MOT beam power, with optimum red detuning of 1.5 $\Gamma$ from the ${^{85}}$Rb cooling transition. \emph{(a)} An average temperature of the vortex and hybrid-MOTs' atom cloud as a function of the MOT beam power. The measured temperature is well below Doppler limit for the vortex-MOT, represented by green dashed line, without any additional sub-Doppler cooling stage. The measured temperature for the hybrid-MOT is slightly higher than the Doppler limit. \emph{(b)} Temperature of the vortex-MOT atom cloud as a function of the of the cooling beam detuning.}
\label{PhiMOT_Temp_New}
\end{figure}

In order to characterize the behaviour of the vortex-MOT, the temperature and the number of trapped atoms are measured as a function of the total MOT beam power and the detuning of the cooling laser from the ${^{85}}$Rb cooling transition. The total MOT beam power consists of the cooling and repump beams in 4.5$\colon$1 ratio respectively, which are equally distributed among four MOT beams at \SI{45}{\degree} to the mirror surface. For the hybrid-MOT the temperature and the atom number are measured only as a function of the total MOT beam power, keeping the  detuning of the cooling laser fixed at 1.5 $\Gamma$ from the ${^{85}}$Rb cooling transition. The hybrid-MOT is very sensitive to alignments and detuning hence only a single configuration is used to compare with the vortex-MOT. The atom number and temperature measured for the vortex-MOT and hybrid-MOT are shown in the Figs., for comparison, \ref{PhiMOT_Atom_New} and \ref{PhiMOT_Temp_New} respectively.


The atom number for both the MOTs are measured by the well known loading rate measurement technique \cite{ArpornthipPressure}. For this measurement, the camera\,(Fig. \ref{experimental_setup}\emph{(b)}) is replaced with a photodiode\,(Thorlabs PDA36A) for the detection of fluorescence signal. The vortex-MOT is loaded for around 3.3 seconds till the atom number is saturated. The mean atom number of 10 such measurements represents a single datapoint in the Fig. \ref{PhiMOT_Atom_New}\emph{(a)}. The maximum atom number of around 6$\times{10^6}$ is observed with a red-detuning of 1.5 $\Gamma$ from the cooling transition and with the maximum available total power of 20 mW. We have repeated the hybrid-MOT atom number measurement same way as the vortex-MOT, but with the optimum detuning of 1.5 $\Gamma$ from the cooling transition and the maximum atom number of around 26$\times{10^6}$ is observed. Further the dependence of the atom number on cooling beam detuning for the vortex-MOT is shown it in the Fig. \ref{PhiMOT_Atom_New}\emph{(b)}. The vortex-MOT cloud size is consistently larger than hybrid-MOT, with average $1/e^{2}$ diameters of approximately 420\,$\mu$m and 380\,$\mu$m respectively. This results in a peak atom number density for the vortex-MOT of $1.5\times 10^{11}$cm$^{-3}$, and $9\times 10^{11}$cm$^{-3}$ for the hybrid-MOT.

The temperature of the atom cloud is measured using the well known time of flight (TOF) measurement. The beams are extinguished to allow the TOF for up to 5\,ms and after the TOF an imaging pulse illuminates the expanded cloud so it can be photographed by a fast CCD camera.  After the image has been taken the cloud is allowed to disperse and then a second image is captured to be used for background subtraction. The TOF is limited by the geometry of the experimental setup and the imaging beam diameter. The width of the atom cloud after each time of flight is determined by a 2D Gaussian fit to the background subtracted image.  The temperature is determined by fitting the Gaussian width to the equation $\sigma^2=\sigma_{0}^2+(k_{\mathrm{B}}T/m)t^2$, where $\sigma_{0}$ and $\sigma$ are the Gaussian widths before and after the expansion (TOF), $k_{\mathrm{B}}$ is the Boltzmann constant, $T$ is the temperature of the cloud, $m$ is the mass of ${^{85}}$Rb and $t$ is the time of flight. Around 50 measurements are taken to determine an average temperature of the cloud. The average temperature of the vortex and hybrid-MOT atom clouds as a function of the MOT beam power is presented in Fig. \ref{PhiMOT_Temp_New}\emph{(a)}. Another plot in Fig. \ref{PhiMOT_Temp_New}\emph{(b)} represents the dependence of the temperature on the cooling beam red-detuning for the vortex-MOT. The vortex-MOT temperature reported here is well below the Doppler limit without any additional sub-Doppler cooling stage, whereas the measured temperature for the hybrid-MOT is slightly above.

\section*{Discussion and Conclusion}
\label{conclusion}
Our results have shown that both the vortex and hybrid-MOT achieves characteristics similar to the conventional mirror-MOTs, however higher atom numbers are obtained with the hybrid-MOT. The vortex-MOT trapped nearly identical number of atoms as the S-MOT, but with lower temperatures as the cooling laser is on all the time. For the hybrid-MOT the balance between the Zeeman and the spatially variable optical force is not at the theoretical optimum. This is due to the difficulty in aligning beams to such high accuracy and also due to imperfect beam intensity balance and nulling of residual magnetic fields. This may have contributed towards the higher temperature, although we believe there may also be optical pumping effects preventing efficient sub-Doppler cooling. What is interesting is the nearly order of magnitude increase of atoms trapped in the hybrid-MOT compared to the vortex-MOT. This cannot be attributed to larger beam overlap volume, nor any other variable which results in greater trap loading. The mechanism for this is still unclear and is under further investigation. For both traps, we did not observe any saturation to the measured atom number with the increased beam power. It should be possible to trap more atoms with more cooling beam power but unfortunately we are limited to the reported maximum beam power.
 
In summary, we have successfully demonstrated two simplified mirror-MOTs with misaligned (displaced) beam geometries which are suitable for use with microfabricated atom chips and any vacuum chambers with restricted optical access. The design is amenable to microfabrication due to the absence of out of plane wires or coils and does not use any frequency selective optics, permitting it to trap different atomic species simultaneously. We have shown the trap parameters closely agree with theoretical predictions and both trap produce comparable number of atoms, as well as temperatures, to standard magneto optical traps. These new MOT designs and techniques will enable to build a portable and miniaturized atom chips for quantum sensors, in addition to finding applications in quantum optics and information.



\section*{Acknowledgements}
The authors would like to thank Mark Bampton for machining work including the ceramic structure used for the MOTs. We would also like to thank Tim Freegarde for lending equipments and optics generously. This work is supported by funding from RAEng, EPSRC, and the UK Quantum Technology Hub for Sensors and Metrology under grant EP/M013294/1.

\section*{Author contributions statement}
R.R. and J.R. made the experimental setup. R.R., A.D. took the data. R.R., J.R., A.D., M.A., M.H. analysed the data and prepared the manuscript. M.H. did the theoretical modelling and supervised all the work. All authors reviewed the manuscript.

\section*{Additional information}
\textbf{Competing financial interests: } The authors declare that they have no competing interests.

\end{document}